\documentclass[referee,sn-aps]{sn-jnl}

\usepackage{graphicx}%
\usepackage{multirow}%
\usepackage{amsmath,amssymb,amsfonts}%
\usepackage{amsthm}%
\usepackage{mathrsfs}%
\usepackage[title]{appendix}%
\usepackage{xcolor}%
\usepackage{textcomp}%
\usepackage{manyfoot}%
\usepackage{booktabs}%
\usepackage{algorithm}%
\usepackage{algorithmicx}%
\usepackage{algpseudocode}%
\usepackage{listings}%
\usepackage{changes}

\usepackage{geometry}
\newif\ifshowdeleted
  \showdeletedfalse   

\let\olddeleted\deleted
\renewcommand{\deleted}[1]{%
  \ifshowdeleted\olddeleted{#1}\else\ignorespaces\fi}

\theoremstyle{thmstyleone}%
\theoremstyle{thmstyletwo}%

\theoremstyle{thmstylethree}%

\title{Mapping academic integrity: global retraction trends explored through a topic lens}

\begin{document}

\title[Mapping Academic Integrity]{Mapping Academic Integrity: Global Retraction Trends Explored through a Topic Lens}

\author[1,2]{\fnm{Zhengyi} \sur{Zhou}}\email{zhouzhengyi@mail.las.ac.cn}
\author[1,2]{\fnm{Ying} \sur{Lou}}\email{louying@mail.las.ac.cn}
\author*[1,2]{\fnm{Zhesi} \sur{Shen}}\email{shenzhs@mail.las.ac.cn}
\author*[1,2]{\fnm{Menghui} \sur{Li}}\email{limh@mail.las.ac.cn}

\affil*[1]{\orgdiv{National Science Library}, \orgname{Chinese Academy of Sciences}, \orgaddress{\city{Beijing}, \postcode{100190}, \country{P. R. China}}}

\affil*[2]{\orgdiv{Department of Information Resources Management, School of Economics and Management}, \orgname{University of Chinese Academy of Sciences}, \orgaddress{\city{Beijing}, \postcode{100190}, \country{P. R. China}}}


\abstract{Scientific publications have long served as the cornerstone of innovation, exhibiting stable growth over the years. Recently, however, retractions have surged dramatically, driven largely by the proliferation of low-quality and fraudulent articles, posing a substantial threat to research integrity. By integrating annual publication and retraction data, this study employs the relative retraction rate ($R^3$) to systematically examine disparities and evolving trends from a topical perspective. Our analysis reveals that the number of retractions has grown significantly faster than that of global publications, yielding an overall retraction rate of $0.12\%$. While retractions occur across all disciplines, substantial disparities exist, ranging from $0.035\%$ in \textit{Physics} to $0.34\%$ in \textit{Computer Science}. This gap widens at finer levels of granularity, reaching roughly $8.99\%$ in \textit{Human-Computer Interaction}. Moreover, unusually high $R^3$ values frequently coincide with rapid publication growth in specific fields. We also developed Retraction Monitor, a web application for monitoring retraction dynamics across diverse fields, enabling stakeholders to visualize these trends and assess risks to research integrity. These findings provide valuable insights for identifying high-risk fields and developing tailored governance policies to strengthen research rigor and mitigate field-specific retraction risks. }

\maketitle

\section{Introduction}\label{Introduction}
Scientific publications constitute the cornerstone of scientific and technological advancement \citep{ASR:Innovation}. They serve not only to archive new theories, methods, and discoveries but also to disseminate findings, thereby establishing a foundation for subsequent studies. Annual publication output reflects the dynamism and intensity of a field \citep{ASR:Innovation,PNAS:Choosing} and may even shape researchers' choice of topics\citep{ASR:Innovation,JOI:tracehot}. However, article quality is equally critical. The proliferation of low-quality or fraudulent publications poses a significant threat, potentially misdirecting scientific inquiry \citep{JAMA:citation}, hindering field development \citep{RIPR:Spread}, wasting resources \citep{eLife:Financialcosts}, and compromising research integrity \citep{PNAS:Misconduct}.

Globally, the volume of scientific publications exhibits a persistent upward trend. Yet the recent surge in retractions has drawn increasing attention from the scientific community \citep{Nature:NEWRECORD,JDIS:Amend}, with the growth rate of retracted articles mow outpacing that of legitimate publications \citep{PNAS:editor}. This trend is partly driven by technological advancements, including plagiarism detection, image duplication analysis, and paper mill identification software, that have enhanced the ability to identify problematic research \citep{Nature:image,Nature:milldetector}. Heightened scrutiny and growing awareness of research integrity have further contributed to this phenomenon, and retractions have become increasingly prevalent across diverse fields \citep{INNOVATION:Sciencemap}.

Most retracted articles are deemed unreliable, whether due to intentional misconduct or unintentional errors \citep{InfectionandImmunity:RetractionIndex}. Their prevalence increases the risk that subsequent articles will cite flawed findings \citep{RIPR:Spread}, potentially triggering a cascade of retractions among unreliable articles \citep{Conference:cascading,InfectionandImmunity:RetractionIndex,JOI:perception}. For example, numerous systematic reviews and meta-analyses have cited retracted publications,  undermining the integrity of evidence-based medicine \citep{BMJ:evidence,JAMA:SR,JAMAMedicine:SR}. This not only erodes the knowledge foundation essential for innovation \citep{JAMA:citation,SCIENTOMETRICS:postcite} but also poses a serious threat to public health \citep{BMJ:MMR,JME:medicalrisk}. A comprehensive understanding of retraction trends is therefore crucial.

To advance such understanding, three key questions merit investigation. First, what is the global prevalence of retractions, and what trends are emerging?  At present, most research focuses on specific entities, including countries \citep{Nature:NEWRECORD,Publications:China}, institutions \citep{arxiv:RI2}, journals \citep{CLINICAL:journal,RIPR:journal,Metrics:socialJournal}, diseases \citep{CLINICAL:journal,BJU:urology,ACCOUNTABILITY:hematology,ACCOUNTABILITY:COVID-19}, and research fields \citep{PNAS:editor,bioRxiv:ncRNA,arxiv:landscape,Metrics:socialJournal}.  While these studies offer valuable insights, they are subject to inherent limitations, primarily stemming from topic heterogeneity \citep{PNAS:editor,bioRxiv:ncRNA,arxiv:landscape,Metrics:socialJournal},  restricted topical coverage, disparities in dataset construction standards \citep{Metrics:socialJournal}, sampling biases \citep{Publications:China}, and inconsistencies in time windows. Consequently, reported retraction rates fluctuate widely, ranging from $2$ to $8$ per $10,000$ publications \citep{InfectionandImmunity:RetractionIndex,bioRxiv:ncRNA}, obscuring the true global picture. Furthermore, large-scale retractions continue to emerge, steadily inflating the global retraction rate. Notably, the annual retraction rate exceeded $20$ per $10,000$ publications in 2022, largely driven by mass retractions from Hindawi \citep{Nature:NEWRECORD}. Given the continuous evolution of retraction patterns, such fragmented studies are insufficient to capture the comprehensive landscape. To  accurately quantify global prevalence, research based on a unified, comprehensive  retraction database is imperative.

Second, are retraction prevalence and trends consistent across different disciplines and their subfields? Evidence suggests that the distribution of retractions is highly heterogeneous \citep{INNOVATION:Sciencemap}. For instance, the retraction rates range from $1.7$ per $10,000$ publications in \textit{Physics} to $17.4$ in \textit{Electrical Engineering, Electronics \& Computer Science} (\textit{EE \& Comp Sci}) \citep{INNOVATION:Sciencemap}. While some studies examine specific diseases or subfields \citep{CLINICAL:journal,BJU:urology,ACCOUNTABILITY:hematology,ACCOUNTABILITY:COVID-19,PNAS:editor,Metrics:socialJournal}, they often rely on isolated case studies that focus narrowly on retraction rates \citep{ACCOUNTABILITY:COVID-19}, underlying causes \citep{CLINICAL:journal}, gender disparity \citep{JOI:gender} or post-retraction citations \citep{SCIENTOMETRICS:postcite}.  This reliance on disparate data standards and time frames hinders direct cross-study comparisons. Moreover, existing research is heavily biased toward \textit{Clinical and Life Sciences} (\textit{Clin \& Life Sci}) owing to their health implications, leaving other domains substantially under-investigated. Notably, although \textit{EE \& Comp Sci} exhibits a higher retraction rate than \textit{Clin \& Life Sci} \citep{INNOVATION:Sciencemap}, it receives significantly less scholarly attention. To address this disciplinary imbalance, it is essential to examine retractions across diverse fields using a unified, comprehensive dataset. A time- and scale-independent indicator is also needed to effectively normalize retraction severity and clarify evolving trends.

Third, what patterns are associated with the escalation of retraction prevalence? Some studies have examined factors linked to systematic scientific fraud \citep{RIPR:journal,PNAS:editor,arxiv:landscape}. For example, open access publishing may be associated with the rise of paper mills, predatory journals, and academic brokers \citep{PNAS:editor}. Additionally, under-investigated topics tend to coincide with higher rates of fraudulent articles \citep{BIOMARKER:fraudulent,arxiv:landscape}, and shortened editorial timelines may co-occur with journals that are targeted by fraudulent submissions \citep{RIPR:journal}. However, these analyses draw on aggregated data from specific fields, limiting their generalizability and practical application. By examining the evolution of retraction trends across diverse topics, we aim to identify patterns correlated with retraction prevalence. 

To address these gaps and empower stakeholders to monitor the evolving landscape of their specific fields, there is a pressing need for a centralized tool that visualizes retraction trends using granular, real-time data. Such a tool would provide valuable insights for strengthening the governance of research integrity.

In response, this study leverages a comprehensive database of over $62,000$ retracted articles, with citation topics serving as the primary unit of analysis. We conduct a detailed examination of publication volume, retraction counts, retraction rates, and their growth trends, tracing their evolutionary patterns from disciplines down to granular subtopics. Our findings indicate that annual retraction numbers have remained consistently elevated in recent years. Furthermore, retractions are widespread, spanning all disciplines and most topics, with certain topics exhibiting markedly higher susceptibility. Notably, retraction patterns vary substantially even within a single topic across different subtopics. We also observed that surges in retractions often coincide with the rapid expansion of a field. To facilitate exploration of these trends, we introduce ``Retraction Monitor", an interactive visualization tool that enables users to track retraction dynamics across disciplines and subtopics. This analysis deepens our understanding of retraction phenomena, providing robust data to support future research and policy development.

\section{Materials and Methods}

\subsection{Retracted Articles}
Retracted articles were collected from the Amend platform \citep{JDIS:Amend}, which aggregates retraction data from official journal websites. This process primarily involves collecting the Digital Object Identifiers (DOIs) of both the retraction notices and the corresponding retracted articles, along with the content of the retraction notices. As of the end of 2025, the Amend database contained over $62,000$ distinct retracted publications in English, consisting mainly of research articles and reviews, while conference abstracts were excluded.

\subsection{Publications and Citation Topics}
Citation topics for individual retracted articles were retrieved on January 6, 2026, by querying their DOIs through the InCites dataset of Clarivate’s Web of Science (WoS), last updated on December 10, 2025. These citation topics are structured within a three-level hierarchical framework comprising 10 macro-topics, 326 meso-topics, and 2,478 micro-topics. Each article is assigned a corresponding topic at each hierarchical level. Of the $62,000$ retracted articles, over $50,000$ were successfully mapped to topics. Additionally, longitudinal data on the annual publication volumes (including articles and reviews) and retracted articles for each topic from 2000 to 2024 were extracted from the InCites dataset. 

\subsection{Retraction Rate}
The retraction rate is a metric used to quantify the frequency of retracted articles within a body of publications \citep{InfectionandImmunity:RetractionIndex}. It is calculated as the number of retracted articles published between years $j$ and $i$, multiplied by $10,000$, and then divided by the total number of publications within the same period. The formula is as follows:
\begin{equation}
    r = \frac{\sum^i_j R_y}{\sum^i_j P_y} \times 10000,
\end{equation}
where $R_y$ denotes the number of retracted articles and $P_y$ represents the number of publications in year $y$.

\subsection{Compound Annual Growth Rate}
The Compound Annual Growth Rate (CAGR) is a standard metric for quantifying the mean annual growth rate of publications over a specified time period \citep{Ocean:CAGR}. It is calculated as: 
\begin{equation}
    \beta=((\frac{P_i}{P_j})^{\frac{1}{i-j}}-1) \times 100,
\end{equation}
where $\beta$ represents the CAGR (in percentage), $P_i$ denotes the number of publications in the final year $i$, and $P_j$ denotes the number of publications in the initial year $j$. To determine whether a field is growing faster than the global average, we define the differential CAGR as:
\begin{equation}
    \Delta \text{CAGR} = \beta^f -\beta^g,
\end{equation}
where $\beta^f$ and $\beta^g$ represent the CAGR of a specific field and global publications, respectively.

Notably, the peak years for publication volume and retraction counts often differ across topics. Given the rapid increase in the overall number of retracted articles starting from 2018, which peaked in 2022, we designate the year with the highest retraction count as the final year ($i$) for each topic. The initial year is set to $i-n$ (where $n=4$ unless specified otherwise), allowing for the calculation of growth rates over a consistent 4-year window. Alternatively, a uniform initial and final year can be applied to all topics for calculating growth and retraction rates.

\section{Results}\label{Results}

\subsection{Global Surge in Retractions}

\begin{figure}[ht]
    \centering
    \includegraphics[width=0.7\linewidth]{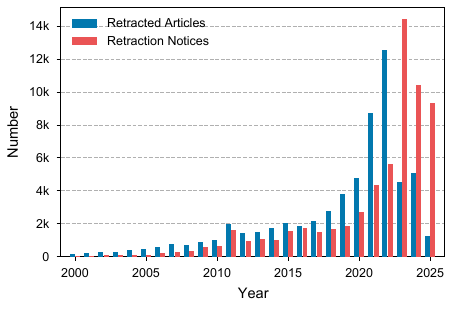}
    \caption{Annual Count of Retracted Articles and Retraction Notices Between 2000 and 2025. }
    \label{fig:retraction_of_year}
\end{figure}

In recent years, the number of retractions has surged dramatically \citep{Nature:NEWRECORD}. The year 2023 set a historical record, with over $14,000$ retraction notices issued.  This trend persisted through 2024 and 2025, which saw over $10,000$ and $9,000$ retractions, respectively. Notably, approximately $12,000$ articles published in 2022 have been retracted, setting a new record for a single publication year (Fig. \ref{fig:retraction_of_year}). The sustained high volume of retractions poses unprecedented challenges for the future governance of research integrity. 

Furthermore, recent years have witnessed a surge in retractions primarily linked to systematic fraud, including fake peer review, AIGC, paper mills, citation manipulation, and other forms of deception. This raises an important question: how many fraudulent articles have yet to be discovered? Strengthening post-publication review processes may help identify potentially fraudulent publications and reduce their impact on subsequent research.

\subsection{Landscape of Retractions}

The above results reveal a surge in retractions in recent years. In particular, the number of retractions has remained high over the past two years (Fig. \ref{fig:retraction_of_year}), and the topical patterns identified in the previous analysis \cite{INNOVATION:Sciencemap} are likely to have shifted considerably. To further explore these patterns, we examined retractions at the macro, meso and micro levels, respectively.

\subsubsection{Macro-Level Disciplinary Disparities}
\begin{figure}[ht]
    \centering
    \includegraphics[width=0.7\linewidth]{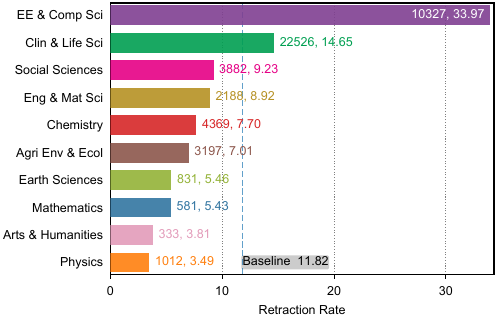}
    \caption{ Retraction Rate by Discipline. This chart illustrates the retraction rates (retracted articles per $10,000$ publications) across various disciplines, along with the total number of retracted articles for each. The dashed line represents the average retraction rate in WoS.}
    \label{fig:retraction_macro}
\end{figure}

According to the Incites database from WoS, between 2000 and 2024, journals indexed by SCI, SSCI, and ESCI published nearly $42$ million articles, of which over $49,000$ were retracted for various reasons. This corresponds to an overall retraction rate of approximately $11.82$ per $10,000$ articles (Fig. \ref{fig:retraction_macro}), which is significantly higher than rates previously reported \cite{INNOVATION:Sciencemap}. 

However, retraction rates vary considerably across macro-topics, ranging from $3.49$ to $33.97$ per $10,000$ publications. For instance, \textit{EE \& Comp Sci} recorded over $10,300$ retractions, yielding a retraction rate of $33.97$ per $10,000$, approximately $10$ times higher than that of \textit{Physics} (Fig. \ref{fig:retraction_macro}). Additionally, \textit{Clin \& Life Sci}, a discipline known for its persistent retraction challenges, accounted for over $22,500$ retracted articles, with a retraction rate of $14.65$ per $10,000$ (Fig. \ref{fig:retraction_macro}). Given their high retraction counts and exceptional retraction rates, these two disciplines have emerged as prominent hotspots for retractions, highlighting a notable disparity in retraction prevalence across fields. Compared with the 2024 findings \cite{INNOVATION:Sciencemap}, both the retraction counts and rates have increased substantially across disciplines.

\subsubsection{Meso-Level Topical Distribution }
\begin{figure}[ht]
    \centering
    \includegraphics{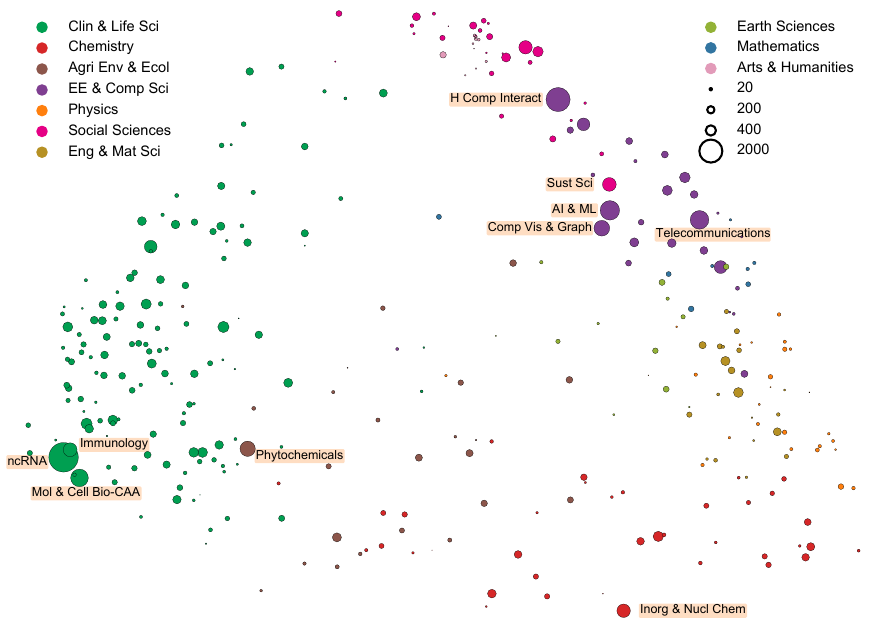}
    \caption{Science Map of Retractions. Each circle represents a meso-topic, sized by the number of retracted articles it contains, with the distance between circles indicating topic similarity based on mutual citations. The top 10 topics are highlighted.}
    \label{fig:science map}
\end{figure}

To investigate this phenomenon at a finer resolution, we turned to the meso-topic level. Empirical evidence confirms that $324$ out of $326$ meso-topics have at least one retracted article. The distribution of retractions was visualized on a science map generated by VOSviewer, which illustrates the citation network among meso-topics (Fig. \ref{fig:science map}). In this map, each circle represents a meso-topic, with its size scaled to the number of retracted articles. The distance between circles reflects topic similarity based on mutual citations \citep{INNOVATION:Sciencemap}.

The number of retractions varies significantly across meso-topics. The highlighted topics in Fig. \ref{fig:science map} represent the top 10 meso-topics by retraction count. These results show some discrepancies compared to the 2024 findings \citep{INNOVATION:Sciencemap}. Notably, \textit{Human-Computer Interaction} (\textit{H Comp Interact}), \textit{Inorganic \& Nuclear Chemistry} (\textit{Inorg \& Nucl Chem}), and \textit{Sustainability Science} (\textit{Sust Sci}) have newly entered the top 10 rankings. Strikingly, \textit{H Comp Interact} ranks second with $2,210$ retracted articles, just behind \textit{Micro \& Long Noncoding RNA}(\textit{ncRNA}) ($3,288$), indicating its emergence as a new retraction hotspot. Similarly, publication volume exhibits considerable heterogeneity across meso-topics (Fig. \ref{fig:sciencemap_pub}). However, among the top 10 most published topics, only \textit{Phytochemicals} also ranked within the top 10 for retractions (Fig. \ref{fig:science map}), suggesting that high publication volume is not inherently associated with a high number of retractions. 

\subsubsection{Meso- and Micro-Level Retraction Severity}
Beyond the top-ranked topics, numerous topics report relatively fewer retractions but also have limited publication volumes, resulting in disproportionately high retraction rates. These fields are also grappling with significant challenges. Thus, relying exclusively on absolute retraction counts or simple rates may not fully capture the true severity of the issue, especially without a global baseline. To accurately assess retraction severity within a specific topic, it is essential to consider both the proportion of retractions and publications against their respective global totals.

\begin{figure}[ht]
    \centering
    \includegraphics{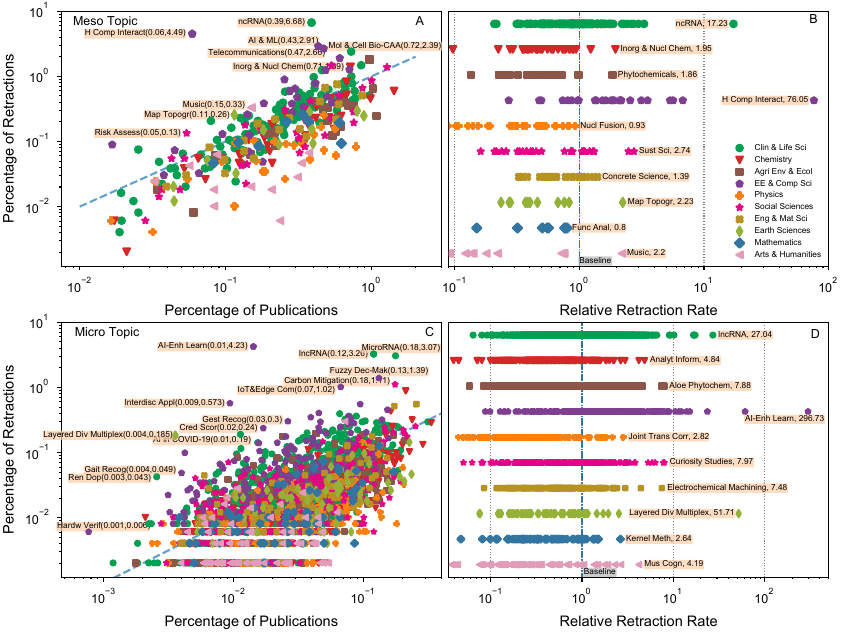}
    \caption{Characterizing Retraction Severity Across Topics. Percentage of retractions relative to publications across various meso-topics (A) and micro-topics (C), with the dashed line indicating the diagonal. Topics where the proportion of retractions significantly exceeds that of publications are highlighted. Relative Retraction Rate $R^3$ across various meso-topics (B) and micro-topics (D), topics with the highest $R^3$ values are highlighted. The dashed line represents the overall retraction severity. $R^3>1$ indicates a severe retraction issue, while $R^3<1$ suggests a milder problem.}
    \label{fig:RI-meso}
\end{figure}

Figure \ref{fig:RI-meso}A compares the percentage of retractions to the percentage of publications across various meso-topics. Several topics exhibit disproportionately high retraction shares relative to their publication shares. For instance,  \textit{ncRNA} and \textit{H Comp Interact} constitute only $0.39\%$ and $0.06\%$ of total publications, yet contribute $6.68\%$ and $4.49\%$ of all retractions, respectively. Many topics in \textit{Clin \& Life Sci} and \textit{EE \& Comp Sci} consistently show retraction percentages that surpass their publication shares, whereas almost all topics in \textit{Mathematics}, \textit{Physics}, and \textit{Chemistry} display the opposite trend. 

To quantify this relationship, we define the Relative Retraction Rate ($R^3$) as the ratio of a topic's share of global retractions to its share of global publications \citep{INNOVATION:Sciencemap}, 
\begin{equation}
    R^3_i=\frac{\frac{R_i}{\sum_j R_j}}{\frac{P_i}{\sum_j P_j}},
\end{equation}
where $P_i$ and $R_i$ denote the number of publications and retracted articles in entity $i$, respectively. $R^3$ provides an effective, scale-independent metric for assessing the severity of retractions across specific entities such as topics, journals, and institutions. Furthermore, annual $R^3$ values enable longitudinal comparisons, offering a comprehensive perspective on retraction trends over time.

Figure \ref{fig:RI-meso}B illustrates the $R^3$ across various meso-topics. The baseline of $1$ represents the global average retraction rate. An $R^3>1$ indicates higher than average retraction severity, while an $R^3<1$ signifies lower than average severity. Notably, \textit{H Comp Interact} exhibits the highest retraction severity, with a peak $R^3$ value of $76.05$, meaning its retraction rate is $76$ times the global average of $11.82$, corresponding to approximately $898.9$ retracted articles per $10,000$ publications. Furthermore, most topics within \textit{EE \& Comp Sci} show $R^3$ values above 1, a pattern also seen in approximately half of the meso-topics in \textit{Clin \& Life Sci}. In contrast, nearly all meso-topics in \textit{Physics} and \textit{Mathematics} exhibit $R^3$ values below $1$.

At the micro-topic level, disciplines beyond \textit{Clin \& Life Sci} and \textit{EE \& Comp Sci} also show retraction shares exceeding their publication proportions (Fig. \ref{fig:RI-meso}C). For example, in \textit{Earth Sciences}, \textit{Layered Div Multiplex} accounts for merely $0.004\%$ of publications but $0.185\%$ of retractions, resulting in an $R^3$ of approximately $51.71$. Similarly, \textit{AI-enhanced learning} (\textit{AI-Enh Learn}) in \textit{H Comp Interact} accounts for $0.01\%$ publications yet contributes $4.23\%$ of retractions, yielding a retraction rate over $296.73$ times the global average. 

Across all disciplines, a substantial number of micro-topics exhibit retraction rates markedly higher than the global average. Notably, even in \textit{Mathematics} and \textit{Physics}, where overall retraction rates are considerably lower, numerous micro-topics still display retraction rates well above the global baseline (Fig. \ref{fig:RI-meso}D). This indicates that a finer-grained categorization is essential for a precise understanding of retraction severity. Aggregate metrics often obscure substantial intra-field heterogeneity, potentially understating the challenges in specific sub-topics. Acknowledging this variation can help identify at-risk areas and guide strategies to curb the dissemination of unreliable articles.

\subsection{Dynamics of Retractions}

Comparison with earlier static findings reveals that the substantial increase in retractions has considerably reshaped the retraction landscape across topics. To better understand these evolving patterns, we now turn to a dynamic analysis of topic-level trends, aiming to identify distinctive characteristics of retraction dynamics over time.

\subsubsection{Macro-Level Growth Trends}
\begin{figure}[ht]
    \centering
    \includegraphics[width=1\linewidth]{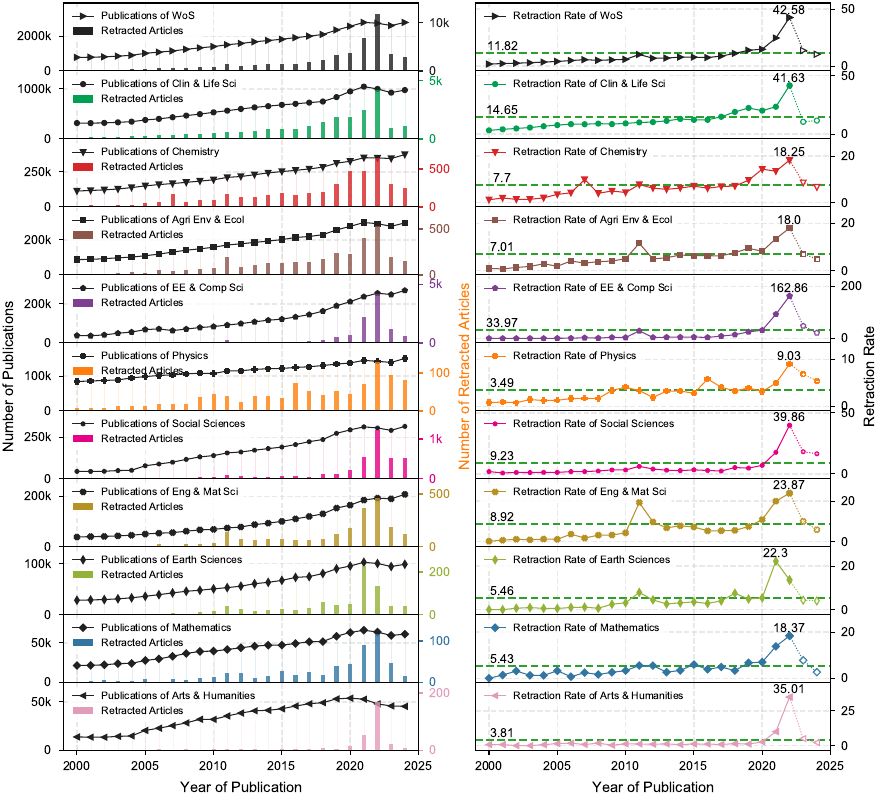}
    \caption{Trends in Publications, Retractions, and Retraction Rates across Macro-topics. Notably, retraction rates for all disciplines declined after 2022. This apparent decrease is likely attributable to the time lag inherent in post-publication review processes, implying that current data are incomplete. Dashed lines represent the disciplinary average retraction rates. Hollow markers indicate incomplete data due to the time lag between publication and retraction. These values are likely underestimated and should be interpreted with caution.}
    \label{fig:evolution_of_macro}
\end{figure}

The total number of publications in WoS grew substantially from $785,000$ in 2000 to $2,800,000$ in 2021, followed by a slight decline, corresponding to a CAGR of $6.24\%$. In contrast, retracted articles surged from $140$ to over $11,700$ by 2022, at a CAGR of $22.29\%$, nearly four times the growth rate of publications. Notably, the retraction rate in 2022 reached $42.58$ per $10,000$, markedly exceeding the average rate of $11.82$ observed from 2000 to 2024 (Fig. \ref{fig:evolution_of_macro}). This indicates that retractions are increasing at a substantially faster pace than the overall scientific enterprise. However, given the time lag inherent in investigating and retracting articles, the current retraction rate for recent years is likely underestimated. Nevertheless, these findings reveal a concerning upward trend in the proportion of compromised research outputs.

Further analysis of individual disciplines indicates that trends in publications, retracted articles, and retraction rates are consistent with global patterns. However, significant discrepancies emerge in peak retraction rates: \textit{Physics} records a maximum retraction rate of $9.03$ per $10,000$, while \textit{EE \& Comp Sci} exhibits a substantially higher peak of $162.86$ (Fig. \ref{fig:evolution_of_macro}). This further underscores that systematic fraud is a pervasive concern across all disciplines.

\begin{figure}[ht]
    \centering
    \includegraphics[width=1\linewidth]{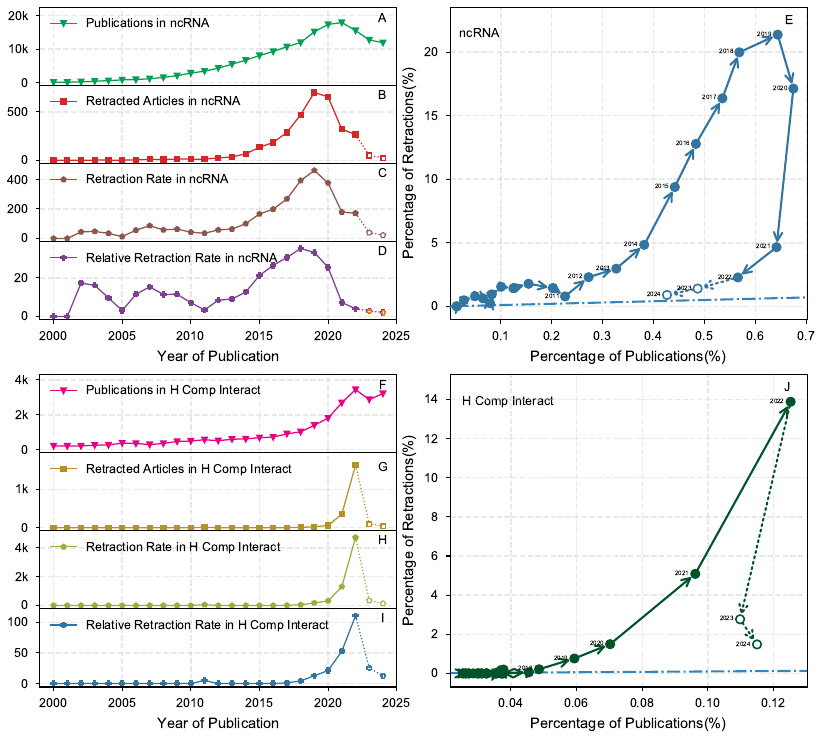}
    \caption{Evolving Retraction Trends at the Meso-Topic Level. Annual trends in publications (A, F), retracted articles (B, G), retraction rates (C, H), relative retraction rates (D, I), and retraction percentage relative to publications (E, J; dot-dashed line indicates the diagonal) for \textit{ncRNA} and \textit{H Comp Interact}.}
    \label{fig:RNA}
\end{figure}

\subsubsection{Meso- and Micro-Level Expansion and Retraction}

The proportion of publications serves as a robust indicator of a topic's expansion relative to global research output, reflecting general trends even as growth rates vary significantly across topics. This is exemplified by \textit{ncRNA} and \textit{H Comp Interact} (Fig. \ref{fig:RNA}). For \textit{ncRNA}, publications surged from $2,840$ (2010) to $15,200$ (2019)  with a CAGR of $20.49\%$, $14$ percentage points above the global average, while retracted articles soared from $12$ to $701$ (CAGR: $57.14\%$) (Fig. \ref{fig:RNA} A and B). As a result, the retraction rate climbed from $42$ to $461$ per $10,000$, reaching an $R^3$ value of $33.20$ in 2019 (Fig. \ref{fig:RNA} C and D). Correspondingly, the publication share increased from $0.20\%$ to $0.64\%$, while the retraction share jumped from $1.44\%$ to $21.37\%$ (Fig. \ref{fig:RNA} E), a pattern that has been linked to paper mills \citep{INNOVATION:Sciencemap,arxiv:landscape}. 

\begin{figure}[ht]
    \centering
    \includegraphics[width=1\linewidth]{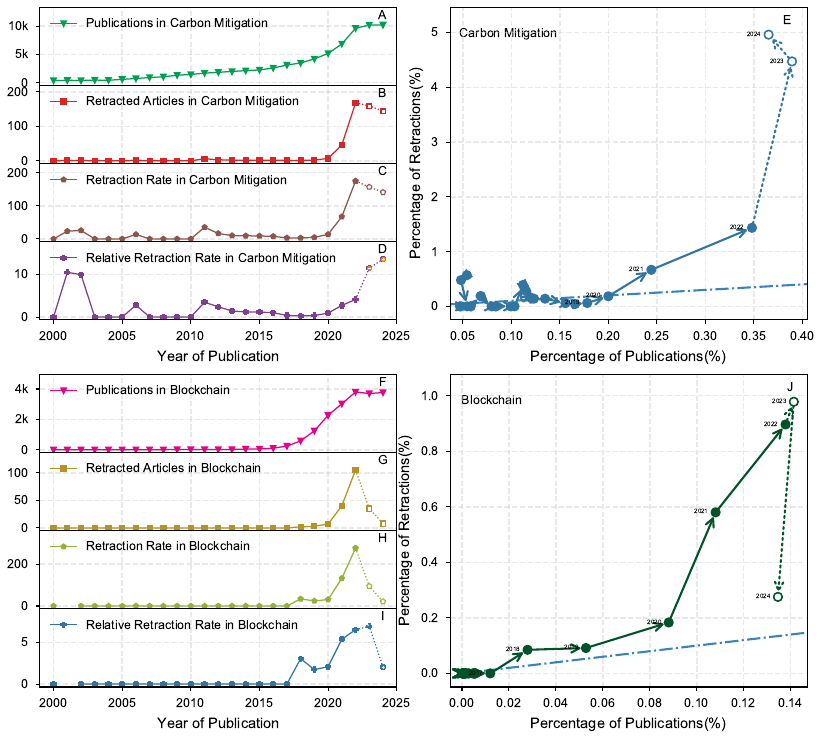}
    \caption{Evolving Retraction Trends at the Micro-Topic Level. Annual trends in publications (A, F), retracted articles (B, G), retraction rates (C, H), relative retraction rates (D, I), and retraction percentage relative to publications (E, J; dot-dashed line indicates the diagonal) for \textit{Carbon Mitigation} and \textit{Blockchain}.}
    \label{fig:Carbon-Blockchain}
\end{figure}

In \textit{H Comp Interact}, articles grew from $1,022$ (2018) to $3,445$ (2022) with a CAGR of $35.50\%$, raising the publication share from $0.05\%$ to $0.13\%$ (Fig. \ref{fig:RNA} F and J). However, retracted articles surged from $5$ to $1,626$ (CAGR: $324.66\%$), increasing their retraction share from $0.21\%$ to $13.88\%$. As a result, the retraction rate hit $4,719.88$ per $10,000$, yielding an $R^3$ of $110.86$ in 2022. Notably, more than $93.66\%$ ($2,070$ of $2,210$) of retracted papers involved cases of fake peer review. Despite a recent decline, $R^3$ remains above $1$, indicating that retractions in these fields continue to pose a significant concern. Collectively, these cases show that the retraction volume is increasing substantially faster than  publication volume.

At the micro-topic level, a recurring pattern emerges in which the surge in retractions is closely accompanied by rapid growth in publication volume (Fig. \ref{fig:Carbon-Blockchain}). For instance, in fields such as \textit{Carbon Reduction} and \textit{Blockchain}, sharp increases in publication output have coincided with marked rises in retraction numbers, resulting in persistently elevated retraction rates. Furthermore, the proportion of retracted articles relative to publication output has grown rapidly. In these fields, rapid expansion is associated with retraction rates that remain consistently above the global average.

\subsubsection{Relationship Between Growth Rate and Retraction Severity}

\begin{figure}[ht]
    \centering
    \includegraphics[width=1\linewidth]{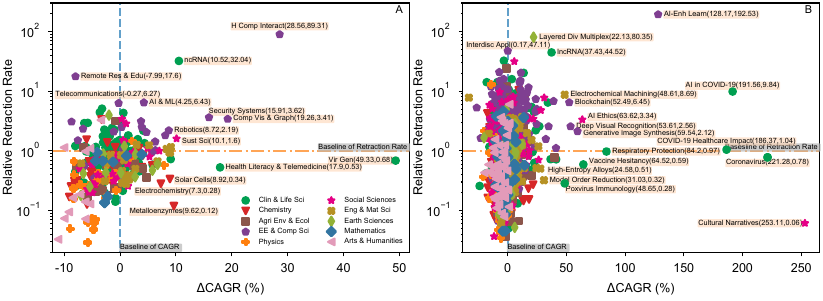}    
    \caption{Relative Retraction Rate $R^3$ in relation to $\Delta$CAGR (disparity between the CAGRs of topics and total publications) for (A) meso-topics and (B) micro-topics. The dashed line represents a growth rate equal to the overall growth rate, while the dash-dotted line indicates a retraction rate consistent with the overall retraction rate. Typical outlier topics are highlighted.}
    \label{fig:CAGR}
\end{figure}

Figure \ref{fig:CAGR} illustrates the relationship between $R^3$ and CAGR across various topics. A cluster of topics exhibits exceptionally high values for both metrics. For instance, \textit{H Comp Interact} shows a $\Delta$CAGR of $28.56$ percentage points above the global average, accompanied by a concerning $R^3$ of $89.31$ (Fig. \ref{fig:CAGR}A). Notably, the actual retraction rate surged to $1,991$ per $10,000$ between 2018 and 2022. This pattern is amplified at the micro-topic level. For example, \textit{AI-Enh Learn} displays a $\Delta$CAGR of $128$ percentage points and an $R^3$ of $192.53$, while \textit{AI in COVID-19} shows a $\Delta$CAGR of $191.56$ percentage points and an $R^3$ of $9.84$ (Fig. \ref{fig:CAGR}B). Collectively, these findings indicate that high retraction prevalence coincides with rapid topic growth.

We recognize that the 4-year window centered on the peak year of retracted articles is a descriptive approach to isolating the most active phase of retraction waves. To ensure this choice does not bias our conclusions, we validated the robustness of the observed patterns using extended time windows (2013--2022 and 2003--2022). As shown in Appendix Figs.  \ref{fig:CAGR_2003} and \ref{fig:CAGR_2013}, the results remained qualitatively consistent. Additionally, controlling for the CAGR time frame reveals that topics with elevated $R^3$ values invariably display high growth rates, suggesting this association  is not an artifact of the specific window selected. 

\subsection{Retraction Monitor: An Interactive Tracking Tool}

To translate the patterns identified in our analysis into actionable insights, we developed "Retraction Monitor," an open-source web application built on Python and Streamlit. The tool enables stakeholders to track retraction dynamics across the WoS topic hierarchy interactively. Users can explore temporal trends in retraction counts, $R^3$ values, CAGR, and other metrics for specific topics at the global, macro, meso, and micro levels. Unlike static charts, the application allows dynamic data filtering by hierarchical topic categories, with visualizations accompanied by the underlying statistical data to ensure full transparency. The responsive web interface requires no programming background, making it accessible to researchers, journal editors, and research administrators. The project is open-sourced under the GPL-2.0 license and publicly available at \url{https://gitee.com/zhesi-shen/retraction-monitor}. A detailed description of the technical implementation and core features is provided in Appendix \ref{tools}.

\section{Discussion}
Under the ``publish or perish" culture, researchers face intense pressure that is closely linked to questionable practices. This manifests either as direct misconduct (e.g., fabrication, plagiarism) or as a strategic shift toward trending topics characterized by high output and unresolved issues \citep{NATURE:hospital, NRC:truth}. These high-visibility fields, however, are particularly vulnerable to systematic fraud. Notably, numerous under-investigated non-coding RNAs have emerged as a prime target of such misconduct \citep{BIOMARKER:fraudulent, arxiv:landscape}, with nearly half of retracted articles in \textit{ncRNA} linked to paper mills \citep{INNOVATION:Sciencemap,bioRxiv:ncRNA}. Indeed, paper mills have extended well beyond \textit{ncRNA}, penetrating fields such as \textit{Artificial Intelligence \& Machine Learning} (\textit{AI \& ML}), \textit{Inorg \& Nucl Chem}, \textit{Phytochemicals}, \textit{Molecular \& Cell Biology}, among others. This suggests a detrimental feedback loop: rising publication volumes may enable large-scale fraud, while fraudulent articles, in turn, may inflate field growth metrics, masking genuine scientific progress. 

Interpreting retraction metrics requires caution due to significant heterogeneity in disciplinary retraction cultures. Criteria for retraction and the willingness to issue them vary substantially across fields. Consequently, a high retraction rate may not straightforwardly indicate poor quality—it may instead reflect a robust community infrastructure for detecting unreliable publications. Conversely, a low $R^3$ might obscure underlying issues in fields lacking formal mechanisms to initiate retractions. Some fields, therefore, may exhibit a high volume of questioned publications alongside a low number of formal retractions, yielding an artificially low $R^3$. A stark example of this phenomenon is observed in the field of \textit{Mesenchymal Stem Cells}. Although \textit{cardiac stem cell} within this field are known to be fraudulent, the anticipated wave of retractions has not occurred, leaving its $R^3$ deceptively low.  While $R^3$ captures only a snapshot of current data, it cannot account for varying attitudes toward retraction. As awareness of scientific integrity deepens, retraction data will likely undergo substantial revision. For instance, $R^3$ in ncRNA and H Comp Interact has experienced rapid growth followed by a swift decline. While this trend may suggest a reduction in systematic fraud, it could also reflect the time lag inherent in the retraction process. It is therefore crucial to distinguish the underlying reasons for these discrepancies to avoid misinterpretation.

Retractions arise from a spectrum of causes, ranging from honest errors to severe misconduct, with academic misconduct accounting for over $80\%$ of cases \citep{JDIS:Amend}. However, current analyses tend to conflate retractions driven by research integrity issues (e.g., data manipulation, paper mills, fake peer review) with those caused by external socio-political pressures or simple honest errors. Although efforts have been made to disentangle these categories, most notably EMBOPress's proposal to distinguish 'withdrawal' from 'retraction' \citep{EMBO:wrong,EMBO:self-correction}, a distinction that has since been discussed \citep{ECI:integrity}, such differentiation remains uncommon in large-scale analyses. Although this aggregation captures the potential risk to research integrity to some extent, it may obscure the true nature of the problem, especially in fields where retractions are primarily due to honest errors. Furthermore, retraction reasons vary significantly across topics. In some fields, fabrication, falsification, and plagiarism are predominant, while in others, systematic fraud is the main driver \citep{INNOVATION:Sciencemap,bioRxiv:ncRNA}. The $R^3$ indicator is more sensitive to systematic fraud (e.g., paper mills, fake peer review, AIGC) but less responsive to traditional fabrication, falsification, and plagiarism. Therefore, the differential impact of retraction reasons must be considered when evaluating the health of a research field.

The persistent and substantial volume of retractions suggests that current measures have not fully identified the problematic research. A significant number of flawed publications remain in the scientific record, as evidenced by the steady stream of articles flagged on post-publication peer review platforms such as PubPeer \citep{PNAS:editor, JASIST:PubPeer}. For example, in the \textit{ncRNA} field, over $9000$ articles have been flagged for concern on PubPeer, yet only about $3000$ have been retracted\citep{bioRxiv:ncRNA}. Monitoring the volume of PubPeer comments per field can serve as a leading indicator for future retraction waves. Therefore, combining post-publication comment data with retraction data may better evaluate the health of fields.

A high $R^3$ value is significantly correlated with a surge in publication volume within a specific field. While rapid field development is not inherently problematic, the co-occurrence of high retraction rates and high growth rates warrants caution. Rapid field growth may be driven by theoretical breakthroughs or an influx of researchers into hot topics. However, the rise in $R^3$ is associated with more complex mechanisms, potentially including technological advances that detect more problematic publications, as well as systematic fraud. Crucially, our data do not allow us to causally disentangle whether the rise in $R^3$ results from increased fraud or from technological advancements in misconduct detection. Although rapid growth does not always correlate with a high $R^3$, explosive expansion in certain fields is concerning. A case in point is the surge in low-quality publications using public health data, which has drawn concern from publishers\citep{Science:publicdata,Nature:publicdata}, leading many to reject database-related submissions\citep{Science:bans}.

\section{Limitation}
Our dataset of retracted articles exclusively includes publications in English, thereby excluding numerous non-English articles from the retraction records. This language bias leads to an underestimation of global retractions. Additionally,  because we used the InCites dataset from WoS to retrieve citation topics, approximately $10,000$ retracted articles lacking assigned topics were excluded, as WoS does not index all scientific publications. As a result, the annual counts of publications and retractions for each topic may be underestimated.

Employing a single metric such as $R^3$ for cross-disciplinary comparison is fundamentally limited by heterogeneity in retraction cultures. Because retraction criteria vary significantly across academic fields, a high $R^3$ in one field (e.g., driven by honest errors) may not indicate the same level of misconduct as a high $R^3$ in another field (e.g., driven by paper mills or fabrication). Furthermore, our analysis aggregates retraction reasons, without distinguish between academic misconduct and honest errors. This conflation obscures the specific drivers underlying retraction trends. 

The current analysis is predominantly based on citation topics and does not incorporate alternative classifications, such as keyword-based searches and AI-based similarity matching between articles. While keyword-based searches in OpenAlex and WoS align with topic-based analyses within the \textit{ncRNA} field \citep{bioRxiv:ncRNA}, the robustness of this approach across other classification systems merits further investigation. A key area for such validation involves employing large language models to identify topical similarities between articles. Furthermore, our analysis is restricted to formal retractions. We explicitly excluded corrections, refutations, and post-publication commentary from platforms such as PubPeer. This limitation stems from the absence of structured, standardized metadata for these sources, which prevents systematic large-scale analysis. Notwithstanding these constraints, we acknowledge that future research leveraging these unstructured sources would undoubtedly yield a more granular understanding of research integrity trends.

Due to the significant time lag inherent in the post-publication review and retraction process, the current number of retractions per topic is likely underestimated, particularly for recent years. This delay implies that the evolutionary trajectory of retraction trends may shift unpredictably as more problematic articles are identified and retracted in the future. Thus, real-time assessment of retraction risks remains  challenging.

\section{Conclusions}
Collectively, our analysis reveals that the escalating prevalence of retractions poses a substantial risk to the scientific community. In recent years, a significant number of articles have been retracted, primarily due to systematic fraud, including paper mills, AIGC, and fake peer review. The growth rate of retracted articles is increasing at a pace that surpasses that of regular publications. While retractions occur across a wide range of topics, the severity of the problem varies across specific domains when academic fields are stratified into micro-topics. Notably, the topics most severely impacted by retractions frequently coincide with those undergoing rapid expansion. To facilitate the understanding of retraction trends across various fields, we have developed a web application,``Retraction Monitor".

These findings underscore two critical implications. First, the true volume of low-quality and fraudulent publications remains uncertain, highlighting the need for collaborative efforts among journals, institutions, and policymakers to scrutinize high-risk domains. Second, while our study identifies priority areas for targeted review, the complete removal of fraudulent content remains a persistent challenge. Future work could move beyond simple retraction counts and leverage granular, field-specific analyses to help safeguard research integrity more effectively.

\section*{Acknowledgments}  
This study is partially supported by the LIS Outstanding Talents Introducing Program, Bureau of Development and Planning, CAS (2022), the Beijing Natural Science Foundation (grant no. 9242006), and the National Natural Science Foundation of China (grant no. 71974017).

\section*{Data availability statement}

The dataset supporting the conclusions of this article is available in the Zenodo repository at \url{https://zenodo.org/doi/10.5281/zenodo.18294748}. 
    
We provide a web application(built with Python \& Streamlit) to visualize the temporal evolution of retractions across topics. The application offers both graphical trends and  datasets for transparency. Access it here: \url{https://gitee.com/zhesi-shen/retraction-monitor}.

\section*{CRediT authorship contribution statement}

Zhengyi Zhou: Data curation, Visualization, Formal analysis, Writing – original draft. Ying Lou: Data curation, Formal analysis, Visualization. Zhesi Shen: Conceptualization, Methodology, Visualization, Writing – review \& editing. Menghui Li: Conceptualization, Formal analysis, Visualization, Writing – review \& editing.

\section*{Disclosure statement} 
The authors report there are no competing interests to declare. 

\section*{Declaration of generative AI use}

During the preparation of this manuscript, the authors used ChatGPT to improve language clarity and readability. The authors reviewed and edited the content as necessary and assume full responsibility for the publication.


\begin{appendices}
\renewcommand{\thefigure}{A\arabic{figure}}
\renewcommand{\thetable}{A\arabic{table}}
\setcounter{figure}{0} 
\setcounter{table}{0} 

\section{Supporting Information}


\subsection{Science Map of Publications}
\begin{figure}[ht]
    \centering
    \includegraphics{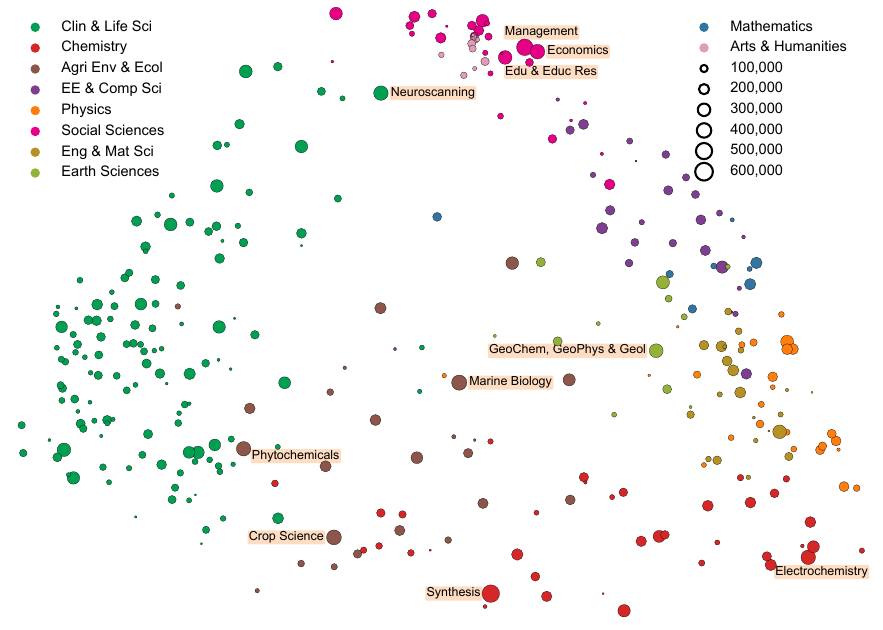}    
    \caption{Science Map of Publications. Each circle represents a meso-topic, sized by the number of publications it contains, with the distance between circles indicating topic similarity based on mutual citations. The top 10 topics are highlighted.}
    \label{fig:sciencemap_pub}
\end{figure}

The science map of publications reveals a significant disparity across meso-topics(Fig. \ref{fig:sciencemap_pub}). For instance, \textit{Synthesis} is the most prolific topic with over $590,000$ publications, whereas the lowest-ranked topic has fewer than $7,000$. However, retraction numbers do not correlate with publication volume. For example, \textit{ncRNA} and \textit{Human Computer Interaction}, with over $160,000$ and $26,000$ publications respectively, contributed to over $3,200$ and $2,200$ retracted articles (Fig. \ref{fig:science map}). In stark contrast, high-volume topics like \textit{Phytochemicals }($400,000+$ publications) and \textit{Synthesis} ($590,000+$ publications) accounted for only $880$ and $290$ retracted articles. This inverse relationship indicates that a high publication volume does not necessarily link to a high number of retractions.

\subsection{Robustness Check: Alternative Time Windows}

\begin{figure}[ht]
    \centering
    \includegraphics{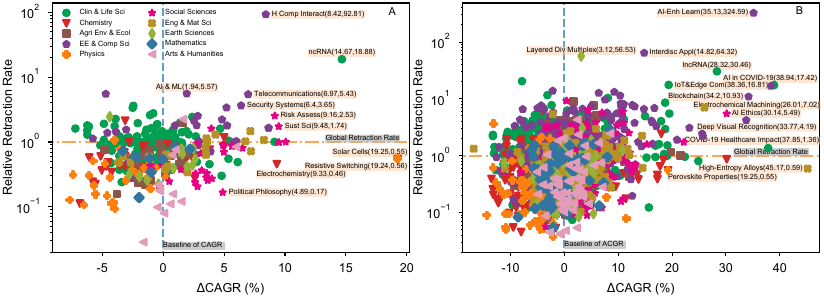}    
    \caption{Relationship between the Relative Retraction Rate $R^3$ and $\Delta$CAGR for (A) meso-topics and (B) micro-topics over the period 2003–2022. The dashed line represents a growth rate that is equal to the overall growth rate, while the dash-dotted line indicates that the retraction rate is consistent with the overall retraction rate. Typical outlier topics have been highlighted.}
    \label{fig:CAGR_2003} 
\end{figure}

\begin{figure}[ht]
    \centering
    \includegraphics{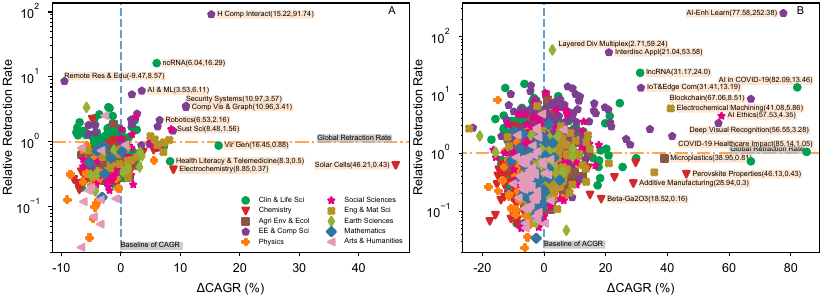}    
    \caption{Relationship between the Relative Retraction Rate $R^3$ and $\Delta$CAGR for (A) meso-topics and (B) micro-topics over the period 2013–2022. The dashed line represents a growth rate that is equal to the overall growth rate, while the dash-dotted line indicates that the retraction rate is consistent with the overall retraction rate. Typical outlier topics have been highlighted.}
    \label{fig:CAGR_2013} 
\end{figure}

To further validate the correlation between high $R^3$ and topic growth rates, we expanded the window for calculating them. For all topics, the terminal year was designated as 2022, the year with the highest global retracted articles, with 2013 and 2003 serving as starting years. Based on these intervals, we calculated the $R^3$ and CAGR for each topic separately (Fig. \ref{fig:CAGR_2003} and \ref{fig:CAGR_2013}).

Our analysis reveals that topics with high $R^3$ consistently exhibit growth rates exceeding the global average. However, the specific values for both growth and retraction rates vary significantly depending on the length of the observation window. In the topic of Human Computer Interaction (H Comp Interact), for instance, CAGRs are $8.4\%$, $15.2\%$, and $28.6\%$ higher than the global maximum growth rate across the respective periods, while the corresponding $R^3$ stands at $92$, $91$, and $89$ (Fig. \ref{fig:CAGR_2003}, \ref{fig:CAGR_2013} and \ref{fig:CAGR}).

\subsection{Retraction Monitor: Technical Implementation and Features}\label{tools}

\textbf{Project Repository}: \url{https://gitee.com/zhesi-shen/retraction-monitor}

\textbf{License}: GPL-2.0

\subsubsection{Overview}

Retraction Monitor is an open-source web application built on Python (version 3.8+) and the Streamlit framework (version 1.28+). It is designed to provide interactive visualization of retraction trends across the Web of Science (WoS) citation topic hierarchy, covering macro-, meso-, and micro-levels. The application addresses the limitations of static charts commonly found in retraction studies by enabling users to dynamically explore data for specific topics of interest.

\subsubsection{Data Source and Processing}

The application draws on the InCites dataset from Clarivate's Web of Science, which provides annual publication and retraction counts for each citation topic from 2000 to 2024. Data are preprocessed and stored in a structured format to enable efficient querying and visualization. Key metrics available for each topic include:
\begin{itemize}
\item Annual publication counts
\item Annual retraction counts
\item Retraction rates (per 10,000 publications)
\item Relative Retraction Rate ($R^3$)
\item Compound Annual Growth Rate (CAGR)
\item $\Delta$CAGR (deviation from the global average CAGR)
\item Publication/Retraction Shares
\end{itemize}

\subsubsection{Core Features}

\begin{itemize}
\item \textbf{Hierarchical Topic Selection}: Users can navigate the three-level WoS topic hierarchy (macro, meso, and micro) via interactive dropdown menus.

\item \textbf{Dynamic Temporal Visualization}: The application generates time-series charts for the selected topic, including annual trends in publications, retracted articles, retraction rates, and $R^3$ values. Charts are rendered using Plotly, providing interactive features such as zooming, panning, and hover tooltips.

\item \textbf{Transparent Data Display}: Alongside each visualization, the underlying statistical data are presented in tabular format. This includes annual retraction counts, percentages, and cumulative metrics, ensuring full reproducibility and allowing users to verify or export the data for further analysis.

\item \textbf{Responsive Interface}: The web interface, powered by Streamlit, requires no programming expertise and is accessible via standard web browsers. 
\end{itemize}

\subsubsection{Application Value and Use Cases}

The tool is intended for a broad audience, including:
\begin{itemize}
\item \textbf{Researchers}: To assess the integrity landscape of specific fields.
\item \textbf{Journal Editors and Publishers}: To monitor retraction trends in fields relevant to their journals and identify emerging integrity challenges.
\item \textbf{Research Administrators and Policymakers}: To inform evidence-based governance strategies by identifying high-risk disciplines.
\item \textbf{Scientometricians and Meta-researchers}: To explore hypotheses related to retraction dynamics and field-level growth patterns.
\end{itemize}

\subsubsection{Deployment and Access}

The application is publicly accessible and can be run locally by cloning the repository and installing the required dependencies listed in the \texttt{requirements.txt} file. Deployment instructions are provided in the repository's \texttt{README.md}. 

\subsubsection{Future Development}

Possible enhancements include: (i) integration of retraction reason categories (e.g., plagiarism, paper mills, fake peer review); (ii) geospatial heatmaps of retraction distributions by country and institution; (iii) predictive modeling of future retraction trends based on historical data; and (iv) incorporation of post-publication peer review data from platforms such as PubPeer to provide a more comprehensive integrity assessment. Community contributions are welcomed under the GPL-2.0 license.

\begin{figure}[ht]
    \centering   
     \includegraphics[width=0.9\linewidth]{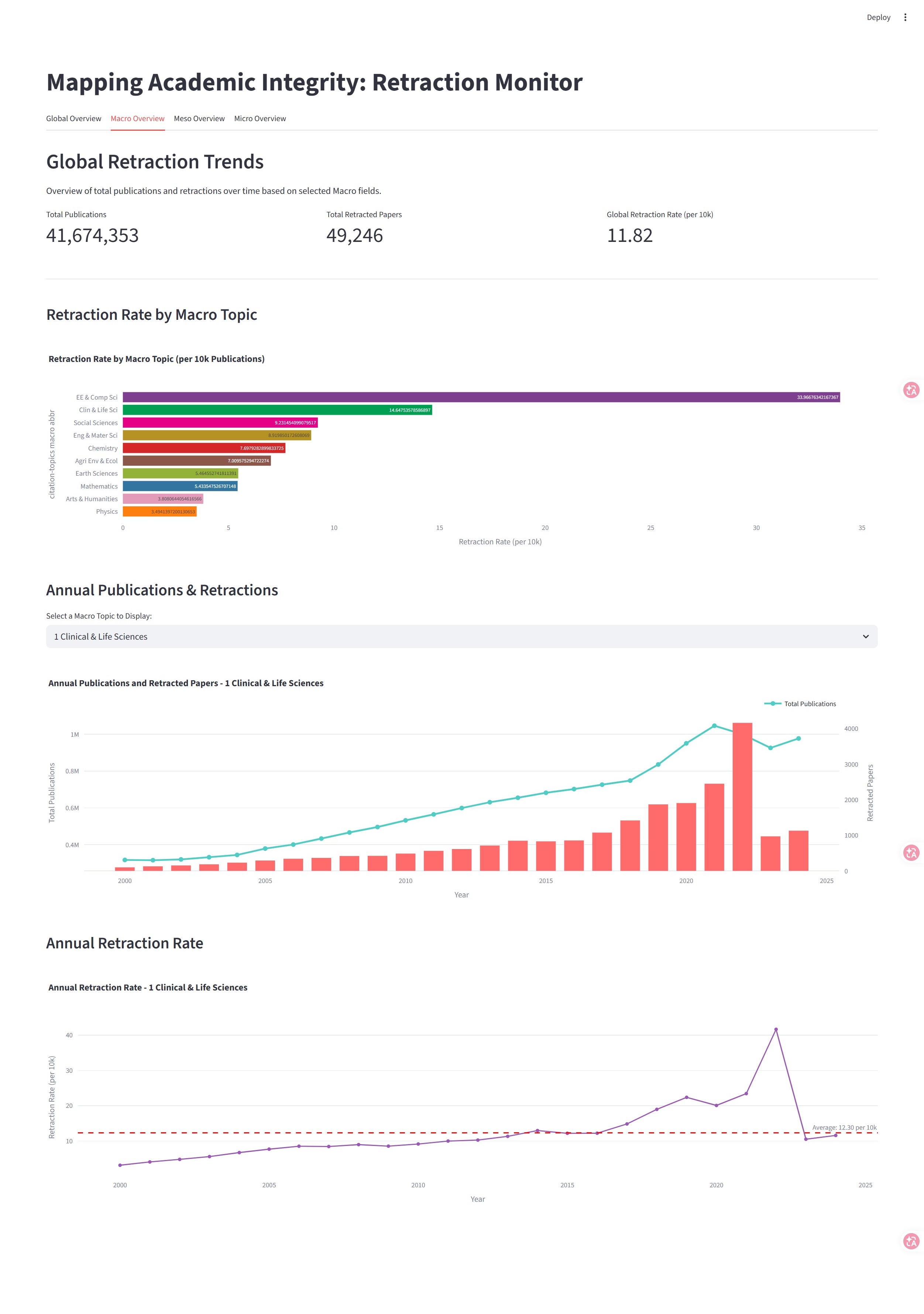}   
    \caption{Application interface of Retraction Monitor.}
    \label{fig:monitor}
\end{figure}

\end{appendices}

\end{document}